\documentclass[aps,prl,twocolumn,showpacs,groupedaddress,superscriptaddress,footinbib,longbibliography]{revtex4-2}
\usepackage{mathrsfs}
\usepackage{natbib,hyperref}
\setcitestyle{square,sort&compress,comma,numbers}
\hypersetup{colorlinks=true, citecolor=blue, urlcolor=blue, linkcolor=blue}
\usepackage[utf8]{inputenc}
\usepackage{graphicx}
\usepackage[normalem]{ulem}
\usepackage[caption=false]{subfig}
\usepackage{amsmath}
\usepackage{amsfonts}
\usepackage{amssymb}
\usepackage{color,soul}

\setulcolor{red}
\usepackage{easyReview}
\usepackage{xcolor}
\usepackage{tikz}
\usepackage{pgfplots}
\usepackage{ulem}
\usepgfplotslibrary{colormaps}
\pgfplotsset{
    colormap={mycolormap}{
        rgb255=(59,76,192)
        rgb255=(255,255,255)
        rgb255=(180,4,38)
    }}

\newcommand\delete[1]{\textcolor{red}{\sout{#1}}}

\long\def\delete#1{}

\begin{document}

\title{Learning microstructure in active matter}

\author{Writu Dasgupta}
\affiliation{Institute for Condensed Matter Physics, Technische Universit{\"a}t Darmstadt, Hochschulstraße 8, 64289 Darmstadt, Germany.}

\author{Suvendu Mandal}
\email{suvendu.mandal@pkm.tu-darmstadt.de}
\affiliation{Institute for Condensed Matter Physics, Technische Universit{\"a}t Darmstadt, Hochschulstraße 8, 64289 Darmstadt, Germany.}

\author{Aritra K. Mukhopadhyay}
\affiliation{Institute for Condensed Matter Physics, Technische Universit{\"a}t Darmstadt, Hochschulstraße 8, 64289 Darmstadt, Germany.}

\author{Benno Liebchen}
\email{benno.liebchen@pkm.tu-darmstadt.de}
\affiliation{Institute for Condensed Matter Physics, Technische Universit{\"a}t Darmstadt, Hochschulstraße 8, 64289 Darmstadt, Germany.}
\begin{abstract}

Understanding microstructure in terms of closed-form expressions is an open challenge in nonequilibrium statistical physics.
We propose a simple and generic method that combines particle-resolved simulations, deep neural networks and symbolic regression to predict the pair-correlation function of passive and active particles. Our analytical closed-form results closely agree with Brownian dynamics simulations, even at relatively large packing fractions and for strong activity. The proposed method is broadly applicable, computationally efficient, and can be used to enhance the predictive power of nonequilibrium continuum theories and for designing pattern formation.
\end{abstract}
\maketitle


The question how macroscopic phenomenon arise from microscopic interactions is central to phenomena across physics, chemistry, and biology, with examples ranging from phase transition ~\cite{hansenPhyRev_1969,kanaiCrystallizationReentrantMelting2015,royallColloidalHardSpheres2024,reisCrystallizationQuasiTwoDimensionalGranular2006} and glass formation~\cite{weeksPropertiesCageRearrangements2002a,banerjeeRoleStructureEntropy2014,nandiRolePairCorrelation2017}, \delete{~\cite{weeksThreeDimensionalDirectImaging2000,royallColloidalHardSpheres2024,gotzeComplexDynamicsGlassForming2008a,boattiniAutonomouslyRevealingHidden2020}} to colloidal self-assembly~\cite{klokkenburgQuantitativeRealSpaceAnalysis2006}, fluid diffusion in porous materials~\cite{bousigeBridgingScalesDisordered2021}, and the self-organization of proteins in the crowded cellular cytosol~\cite{vonbulowDynamicClusterFormation2019}.
In equilibrium systems, the radial distribution function (RDF), $\mathrm{g}(r)$, describes the particle density around a central test particle as a function of distance $r$. Besides characterizing structure, $\mathrm{g}(r)$ plays a pivotal role in relating microscopic structure to macroscopic thermodynamic properties, 
e.g., via the virial and the energy equation ~\cite{hansen_theory_2013,pihlajamaaComparisonIntegralEquation2024,stonesModelFreeMeasurementPair2019}.
Beyond that, $\mathrm{g}(r)$ serves as a key ingredient to understand the collective dynamics of dense and supercooled liquids~\cite{banerjeeRoleStructureEntropy2014,nandiRolePairCorrelation2017}, with subtle variations in its oscillatory shape indicating transitions such as re-entrant glass transition in colloidal-polymer mixtures~\cite{phamMultipleGlassyStates2002} or dynamic slowdown due to confinements~\cite{mandalMultipleReentrantGlass2014}. Thus, $\mathrm{g}(r)$ is not merely a geometric descriptor but a predictive function, linking microscopic structure to macroscopic thermodynamics and collective dynamics, which is essential for understanding complex material properties.

Recently, microstructure-informed theoretical frameworks have been extended to nonequilibrium systems such as active matter, featuring a continuous energy input. Active systems, including self-propelled colloids~\cite{kurzthalerProbingSpatiotemporalDynamics2018,palacciLivingCrystalsLightActivated2013,ginot_nonequilibrium_2015,bechingerActiveParticlesComplex2016,golestanianDesigningPhoreticMicro2007,palacciSedimentationEffectiveTemperature2010,scagliariniUnravellingRolePhoretic2020,garcia-millanOptimalClosedLoopControl2025,zottlEmergentBehaviorActive2016,thutupalliFlowinducedPhaseSeparation2018,fernandez-rodriguezFeedbackcontrolledActiveBrownian2020,paxtonCatalyticNanomotorsAutonomous2004,howseSelfMotileColloidalParticles2007,grauerActiveDroploids2021,ginot_nonequilibrium_2015}, bacterial colonies~\cite{zhang_collective_2010,wensinkMesoscaleTurbulenceLiving2012,peruaniCollectiveMotionNonequilibrium2012,gonzalezlacorteMorphogenesisBacterialCables2025,youGeometryMechanicsMicrodomains2018,dharSelfregulationPhenotypicNoise2022,faluwekiActiveSpaghettiCollective2023,yamanEmergenceActiveNematics2019,curatoloCooperativePatternFormation2020,guillamatTamingActiveTurbulence2017}, and active filaments~\cite{kruseActivelyContractingBundles2000,mandalCrowdingEnhancedDiffusionExact2020,schallerPolarPatternsDriven2010,lemmaActiveMicrophaseSeparation2022,sanchezCiliaLikeBeatingActive2011,serraDefectmediatedDynamicsCoherent2023,sinaasappelParticleSweepingCollection2026}, exhibit collective phenomena such as motility-induced phase separation (MIPS)~\cite{cates_motility_induced_2015,omar_mechanical_2023,speckEffectiveCahnHilliardEquation2014,stenhammarContinuumTheoryPhase2013,filyAthermalPhaseSeparation2012,mandalMotilityInducedTemperatureDifference2019,wittmannActiveBrownianParticles2016,demacedobiniossekStaticStructureActive2018,digregorioFullPhaseDiagram2018,caporussoDynamicsMotilityInducedClusters2023}, flocking~\cite{liebchen_collective_2017,capriniFlockingAlignmentInteractions2023,kreienkampSynchronizationExceptionalPoints2025,kreienkampNonreciprocalAlignmentInduces2024}, and anomalous rheology~\cite{toner_physics_2024}. Central to these behaviors is the pair correlation function $\mathrm{g}(r, \theta)$, which captures both spatial ($r$) and orientational ($\theta$) correlations that can arise from self-propulsion. For instance, the emergence of an orientational asymmetry in the RDF induces MIPS \cite{bialke_microscopic_2013}, explains flocking by turning away~\cite{dasFlockingTurningAway2024}, and
plays a crucial role in quantifying active stresses exerted on passive probe particles in active baths~\cite{paul_force_2022}. The anisotropic RDF also enables predictions of MIPS breakdown in anisotropic systems, as well as the emergence and coexistence of polar and nematic order~\cite{grossmannParticlefieldApproachBridges2020}.

Contrasting its fundamental importance, determining $\mathrm{g}(r)$ and $\mathrm{g}(r,\theta)$ is often challenging. In equilibrium, both for 
simple and complex fluids, classical Density Functional Theory (DFT) provides a powerful tool for deriving 
$\mathrm{g}(r)$ from a free energy functional $\mathcal{F}[\rho]$ ~\cite{dijkmanLearningNeuralFreeEnergy2025,sammullerDeterminingChemicalPotential2025}. However, such functionals are exactly known only for very few cases (in equilibrium)~\cite{hansen_theory_2013}, and generalizations to predict the structure of active systems via dynamical density functional theory are often unreliable far from equilibrium.
While recent machine-learning approaches can determine remarkable representations of $\mathcal{F}[\rho]$ from data~\cite{dijkmanLearningNeuralFreeEnergy2025,sammullerNeuralFunctionalTheory2023,simonMachineLearningApproaches2024,kampaMetadensityFunctionalTheory2025}, they also remain rooted in a (near-)equilibrium framework. 

Accordingly, for active matter, we are currently lacking a general method to predict $\mathrm{g}(r,\theta)$, in particular, in terms of closed-form expressions that are required for the development of continuum theories.
Currently, pioneering existing 
works 
either (i) use linearized Dean-equation approaches that offer analytical expressions in the dilute limit but break down at higher densities~\cite{poncet_pair_2021}, where the full anisotropic structure becomes essential, (ii) 
angularly average $\mathrm{g}(r,\theta)$, erasing anisotropy~\cite{farage_effective_2015}, or (iii) rely on computational approaches~\cite{bialke_microscopic_2013,solon2015pressure,grossmannParticlefieldApproachBridges2020,brokerCollectiveDynamicsPairdistribution2024}.

To address the gap in our understanding of $\mathrm{g}(r,\theta)$, we introduce a simple and generic method that learns (anisotropic) structure from simulations and translates them into interpretable closed-form expressions, that contain the full dependence on system parameters. 
These results can be used in the future to develop analytical theories predicting collective behavior in active matter beyond the low density regime.
\begin{figure*}[tp]
\includegraphics[width=\linewidth]{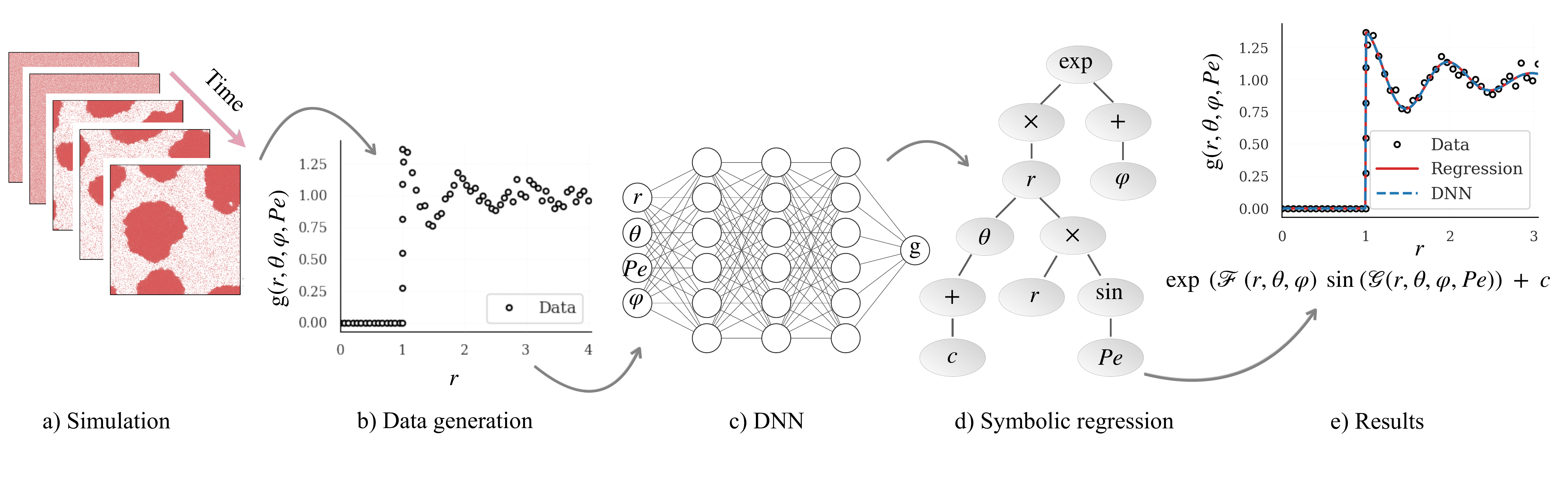}
\caption{\textbf{Schematic illustration of the proposed method.}
(a) Brownian dynamics simulations of active Brownian particles as a function of time, packing fraction $\varphi$, and P\'eclet number $\mathrm{Pe}$. (b) From these snapshots, we compute the radial distribution function, either isotropic $\mathrm{g}(r)$ (passive or angle-averaged) or fully anisotropic $\mathrm{g}(r, \theta)$ (active). 
(c) A deep neural network learns the mapping $(r, \theta, \varphi, \mathrm{Pe}) \mapsto \mathrm{g}(r, \theta)$, providing a smooth, differentiable surrogate for the simulation-measured microstructure. 
(d) Symbolic regression converts the learned surrogate into compact, closed-form analytical expressions. (e) These analytical formulas accurately reproduce near-contact peaks, coordination-shell oscillations, and activity-induced microstructure, offering ready-to-use structural input for nonequilibrium theory.}
\label{fig:method}
\end{figure*}
\paragraph{Model.}
We consider a two-dimensional system of $N=8 \times 10^4$ overdamped active or passive Brownian particles (ABPs or PBPs) and denote the position and orientation of the $i$-th particle by $\mathbf{r}_i$ and $\theta_i$, respectively. Each particle has a diameter $\sigma$, self-propels with velocity $v_0$, and has a translational diffusion coefficient $D_t$. To satisfy the fluctuation-dissipation relation in the equilibrium limit for Newtonian solvents, we fix the rotational diffusion coefficient to $D_r = 3D_t/\sigma^2$. The particles interact via a purely repulsive Weeks--Chandler--Andersen (WCA) potential $U_{\mathrm{WCA}}(r)$, defined as $4\epsilon[(\sigma/r)^{12} - (\sigma/r)^6] + \epsilon$ for $r < 2^{1/6}\sigma$ and zero otherwise, where $r$ is the interparticle distance and $\epsilon$ defines the interaction strength. Subsequently, we non-dimensionalize the system by choosing the length unit $r_u = \sigma$ and the time unit $t_u = \sigma^2/D_t$ (three times the persistence time $1/D_r$). This leads to $\mathbf{r}^* = \mathbf{r}/r_u$ and $t^* = t/t_u$. In these units, the equations of motion are
\begin{eqnarray}
    \dot{\mathbf{r}}^*_i &=& \text{Pe}\ \mathbf{p}_i + \mathbf{F}_i^{\mathrm{int}*} + \sqrt{2}\,\boldsymbol{\xi}^*_i(t^*), \label{eq:eom_pos_dimless} \\
    \dot{\theta}_i &=& \sqrt{6}\,\eta^*_i(t^*), \label{eq:eom_angle_dimless}
\end{eqnarray}
where $\mathrm{Pe} = v_0 \sigma / D_t$ is the P\'eclet number, $\mathbf{p}_i=(\cos\theta_i,\sin\theta_i)$, and $\mathbf{F}_i^{\mathrm{int}*}$ is the dimensionless interaction force. The Gaussian noises $\boldsymbol{\xi}^*$ and $\eta^*$ have zero mean and unit variance. We integrate Eqs.~\eqref{eq:eom_pos_dimless}--\eqref{eq:eom_angle_dimless} using LAMMPS~\cite{LAMMPS} with time step $\Delta t^* = 5 \times 10^{-5}$ in a square domain of side length $L^* = 256$ with periodic boundary conditions. The control parameters are $\mathrm{Pe}$ and the packing fraction $\varphi = N\pi/(4 L^{*2})$ and $\varepsilon^*$ whose precise value is rather unimportant for the emerging collective behavior. We generate a dataset for $\varphi \in [0.20,0.50]$ and $\mathrm{Pe}\in[5,45]$ (Pe = 0: in the equilibrium case) at fixed $\epsilon^*=256$. For each, $(\varphi,\mathrm{Pe})$ pair, we extract 20 statistically independent snapshots from our simulations. From these configurations, we compute $\mathrm{g}(r^*,\theta)$ with radial and angular resolution of $\Delta r^*=0.025$ and $\Delta\theta=4^\circ$.

\paragraph{Deep learning framework.}
To determine $\mathrm{g}(r,\theta)$, we now describe our learning approach, which we later exploit to create a dense dataset as required for the construction of an analytical closed-form expression for $\mathrm{g}(r,\theta)$.

We use the mentioned 20 snapshots for each ($\varphi, \mathrm{Pe}$) combination to train a feed-forward deep neural network (DNN) as a surrogate model for predicting $\mathrm{g}(r, \theta)$ in both active and passive systems. The network takes as input the features $(r, \theta, \varphi, \mathrm{Pe})$ and outputs $\mathrm{g}(r, \theta, \varphi, \mathrm{Pe})$ (see Fig.~\ref{fig:method}). For isotropic systems, the angular coordinate $\theta$ is omitted. Training is carried out using the AdamW optimizer~\cite{loshchilov2017adamw} with a learning rate of $5 \times 10^{-4}$ for 100 epochs (see Supplemental Material (SM) for details of DNN architecture, learning, and loss functions). The DNN achieves root mean square error (RMSE) of $10^{-2}$ for passive systems and between $10^{-2}$ and $10^{-1}$ for active systems (see SM for details).

\par Following DNN training, we apply symbolic regression to the DNN predictions, allowing for continuous input data across area fractions and P\'eclet numbers. We perform symbolic regression by evolving populations of mathematical expressions to minimize a loss function penalized by expression complexity \cite{cranmer2023interpretablemachinelearningscience} (see SM for details).


\paragraph{Equilibrium microstructure from data.}

\begin{figure}[t]
    \centering
    \includegraphics[width=\linewidth]{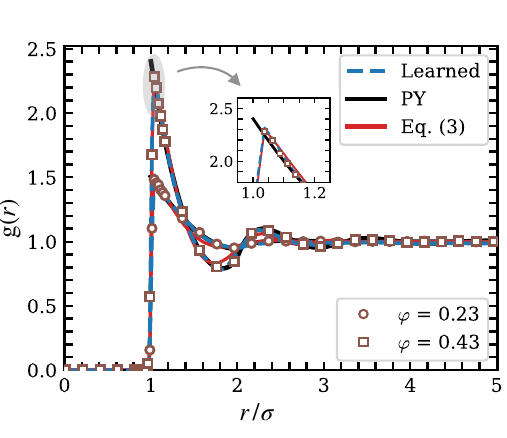}
    \caption{\textbf{Equilibrium microstructure learned from data.}
    Radial distribution function $\mathrm{g}(r)$ of passive Brownian particles at two non-trained packing fractions, $\varphi=0.23$ and $\varphi=0.43$. Symbols represent Brownian dynamics simulation data, the blue dashed line shows predictions from the trained deep neural network (learned), and the black solid line corresponds to the Percus–Yevick (PY) reference solution. Red solid lines represent analytical predictions from Eq.~(\ref{eq:pbp_symbolic}).}
    \label{fig:passive}
\end{figure}

To test our approach, we first explore $\mathrm{g}(r)$ in equilibrium for (almost) hard disks, realized via a steeply repulsive WCA potential~\cite{bollingerHowLocalAverage2015}. For such systems, the Percus–Yevick (PY) closure of the Ornstein–Zernike (OZ) equation provides accurate predictions of $\mathrm{g}(r)$ \cite{percusAnalysisClassicalStatistical1958,hansen_theory_2013}(e.g. analytical Wertheim solution in 3D~\cite{wertheimExactSolutionPercusYevick1963}; semi-analytical solution in 2D~\cite{adda-bediaSolutionPercusYevickEquation2008}.) We use the 2D solution as a benchmark of our learning approach. We now predict $\mathrm{g}($r$)$ directly from a relatively small number of simulations and ask: Can a neural network generalize the structural trends of $\mathrm{g}(r)$ smoothly across varying area fractions and make predictions beyond the trained data? 

Figure~\ref{fig:passive} exhibits this result. The DNN was trained on a range of area fractions $\varphi = 0.2, 0.25, 0.30, 0.35, 0.40, 0.45, 0.50$ and successfully extended its predictions to non-trained area fractions, e.g., at $\varphi = 0.23$ and $\varphi = 0.43$ for which we determine $\mathrm{g}(r)$ from simulations as test cases. For instance, at $\varphi = 0.23$, the DNN captures the characteristic features of a moderately dense fluid, i.e., an initial near-contact peak followed by weak oscillations. As the area fraction increases to $\varphi = 0.43$, the first peak value increases, and subsequent oscillations intensify, signaling enhanced medium-range order. Remarkably, the predicted $\mathrm{g}(r)$ captures these features quantitatively, matching the PY solution and reproducing subtle details such as changes in peak widths and trough depths. The low root mean square error (RMSE $\lesssim 0.03$ across area fractions [see SM]) confirms that the DNN has learned structural principles, not just memorized specific data points.

Having established that the DNN can reliably learn equilibrium structure, we now ask: Can we translate the learned mapping $(r,\varphi) \mapsto \mathrm{g}(r,\varphi)$ into a useful analytical expression? To explore this, we apply symbolic regression to the DNN’s predictions, generating dense datasets across $\varphi=0.2$ to $\varphi=0.5$, yielding a closed-form expression for $\mathrm{g}(r)$:
\begin{align}
\mathrm{g}(r) =\;
& \exp\!\left[
    k_1(k_2\varphi)^{\,r}\,
    \sin\!\big(r^2(\varphi+1)\big)
\right] \nonumber\\[4pt]
&\times
\cos^{k_3}\!\left(
    \exp\!\left[
        k_4 r^{k_5}\right]
\right)
\label{eq:pbp_symbolic}
\end{align}

where $k_i$ with $i \in [1,2,\ldots,5]$ are constants (see SM). While this result is much simpler than known semi-analytical results in 2D~\cite{adda-bediaSolutionPercusYevickEquation2008} and celebrated 3D results~\cite{wertheimExactSolutionPercusYevick1963}, it accurately captures the near-contact peak, coordination-shell oscillations, and their attenuation [see Fig.~\ref{fig:passive}]. The expression also recovers the correct low-density limit $\mathrm{g}(r) \to 1$ as $\varphi \to 0$ and remains in good agreement with simulation results at low packing fractions ($0 \le \varphi < 0.2$) (see SM), despite the limitation of our training data for $\varphi \ge 0.2$.

\paragraph{Active systems: Radial structure $\mathrm{g}(r)$.}
\begin{figure}[t]
    \centering
    \includegraphics[width=\linewidth]{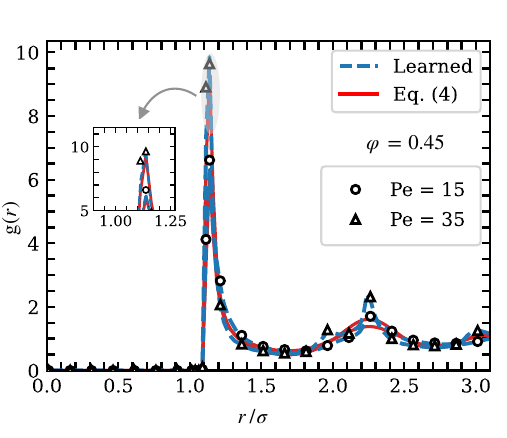}
    \caption{\textbf{Activity-induced angle-averaged microstructure.}
    Validation of analytical predictions from Eq.~(\ref{eq:abp_symbolic}) for various P\'eclet numbers at a fixed area fraction $\varphi=0.45$. Solid lines represent analytical predictions, dashed lines represent learned results from the trained deep neural network, and symbols denote simulation data.
    }
    \label{fig:grfig}
\end{figure}

Unlike equilibrium systems, active matter lacks a unified theoretical framework (such as the minimization of the free energy functional) to obtain $\mathrm{g}(r)$, making them a challenging case for theory. In addition, active particles feature additional orientational degree of freedom (self-propulsion direction) and activity parameters (P\'eclet number). We now ask: How effective is the combination of DNN and symbolic regression to predict the microstructure of active Brownian particles in terms of $\mathrm{g}(r)$?


\begin{figure*}[tp]
\includegraphics[width=0.8\linewidth]{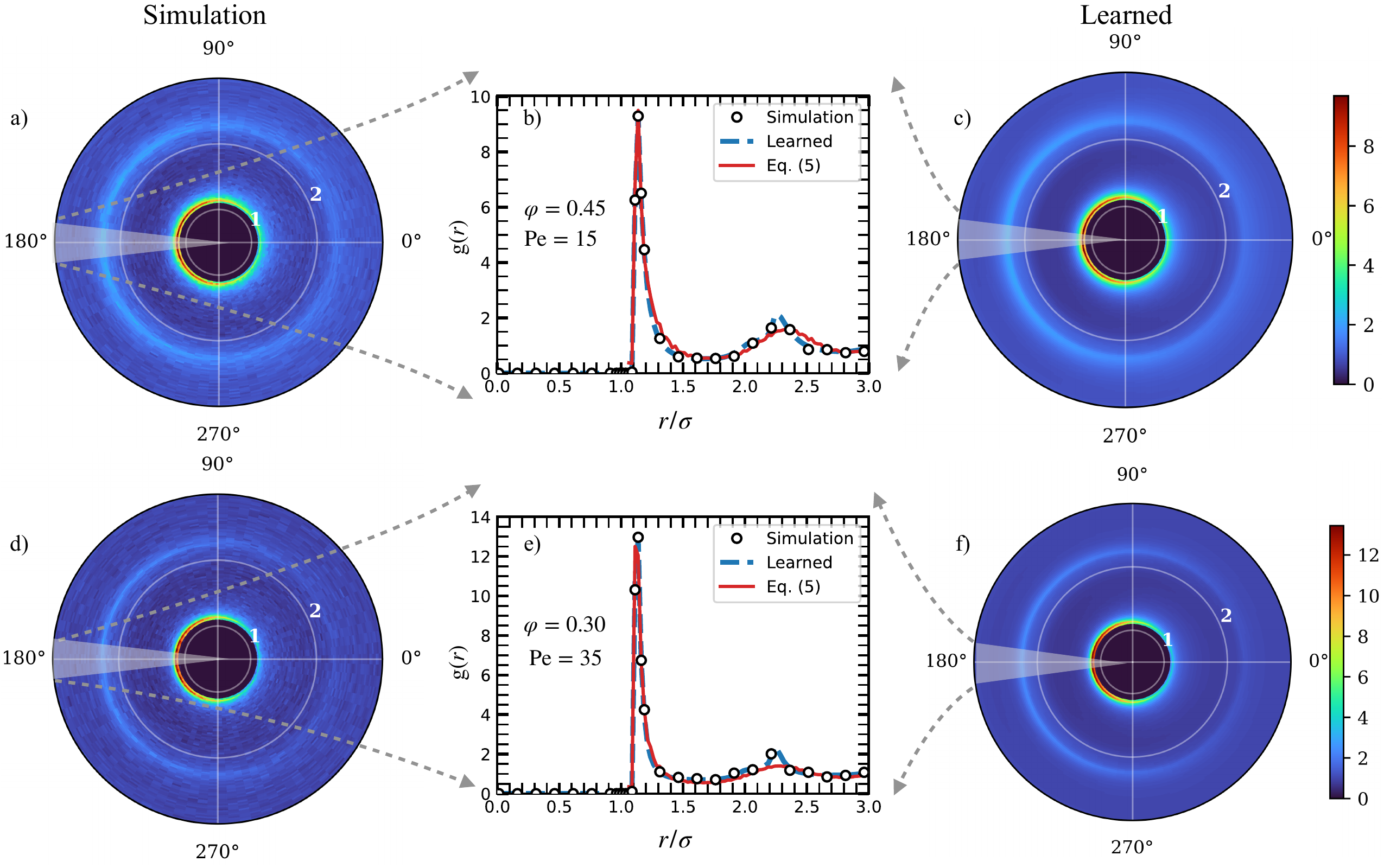}
    \caption{ \label{fig:grthetafig} \textbf{Anisotropic microstructure of active Brownian particles.} Angle-resolved pair correlation function $\mathrm{g}(r,\theta)$ of active Brownian particles (ABPs) obtained from Brownian-dynamics simulations [(a), (d)] for $(\varphi,\mathrm{Pe})=(0.45,15)$ and $(\varphi,\mathrm{Pe})=(0.30,35)$, respectively, and from deep neural network (learned) predictions [(c),(f)] for the same state points. Both simulations and learned predictions reveal a pronounced anisotropic microstructure, characterized by particle accumulation in front of a reference active particle ($\theta=180^\circ$) and depletion in its wake ($\theta=0^\circ$). The central panels [(b), (e)] show radial cuts along the propulsion direction ($\theta=180^\circ$), demonstrating quantitative agreement between simulation data, learned predictions, and the analytical predictions obtained from Eq.~(\ref{eq:abp_angle_symbolic}) over the full radial range.}
\end{figure*}

Figure~\ref{fig:grfig} exemplarily shows the learned microstructure for two non-trained state points. For $\varphi=0.3$ and $\mathrm{Pe}=15$, $\mathrm{g}(r)$ exhibits a near-contact peak and damped oscillations qualitatively similar to equilibrium systems. At higher activity, $\mathrm{Pe}=45$, the near-contact peak becomes more pronounced (see inset of Fig.~\ref{fig:grfig}), and oscillations shift in amplitude and spacing, reflecting the competitive dynamics between activity and steric repulsion. The DNN captures this behavior across all considered area fractions and P\'eclet numbers (see SM). Symbolic regression then converts the learned radial dependence into a compact representation:
\begin{equation}
\mathrm{g}(r) =
\sqrt{
\exp\!\left[
    \mathcal{A}(r,\varphi,\mathrm{Pe})\;
    \mathcal{B}(r,\varphi,\mathrm{Pe})
\right]
} ,
\label{eq:abp_symbolic}
\end{equation}
where $\mathcal{A}$ and $\mathcal{B}$ are relatively simple nonlinear functions (see SM). Eq.~(\ref{eq:abp_symbolic}) offers a nonequilibrium prediction for $\mathrm{g}(r)$, that works even at relatively large $\mathrm{Pe}$, $\varphi$. This result encapsulates how the packing fraction sets the baseline coordination-shell structure, while activity amplifies near-contact correlations and reshapes the oscillatory decay (see Fig.~\ref{fig:grfig}).

\paragraph{Active systems: Anisotropic structure $\mathrm{g}(r, \theta)$.}
While the angle-averaged $\mathrm{g}(r)$ captures how activity modifies the average packing of particles, a defining characteristic of active matter lies in its directional nature~\cite{vrugtWhatExactlyActive2025}. The angle-resolved correlation $\mathrm{g}(r, \theta)$ reveals this directionality by conditioning neighbor statistics along the propulsion axis of a reference particle. In active systems, particles ``push" into the surrounding medium, accumulating neighbors in the direction of motion, while leaving a depleted wake behind (see Fig.~\ref{fig:grthetafig}). This asymmetry plays a critical role, e.g., in the theoretical framework for active stresses~\cite{paul_force_2022} and MIPS~\cite{bialke_microscopic_2013}.

Figure~\ref{fig:grthetafig} compares learned results for $\mathrm{g}(r,\theta)$ with results from Brownian dynamics simulations. Heatmaps illustrate the characteristic accumulation of particles at $\theta = 180^\circ$ (the direction of propulsion) and depletion at $\theta = 0^\circ$ (the rear). As both packing fraction $ \varphi$ and P\'eclet number $\mathrm{Pe}$ increase, the anisotropy becomes more pronounced, signaling the onset of a stronger ``blocking mechanism", leading to the slowdown of particles in regions of enhanced density, which is at the heart of the emergence of MIPS ~\cite{bialke_microscopic_2013, cates_motility_induced_2015}. Also, here, the DNN not only reproduces the qualitative trends but also accurately captures the radial localization of anisotropy near contact, which gradually weakens at larger separations.

Using the quasi-continuous dataset available from the DNN, symbolic regression is employed to construct a compact, analytical form for $\mathrm{g}(r, \theta)$. The resulting expression is:

\begin{equation}
\mathrm{g}(r,\theta) =
r\exp\!\left[
    \mathcal{F}(r,\theta,\varphi)\;
    \sin\mathcal{G}(r,\theta,\varphi,\mathrm{Pe})-c_0
\right]
+ c_1 ,
\label{eq:abp_angle_symbolic}
\end{equation}
where $\mathcal{F}$ and $\mathcal{G}$ are nonlinear 
functions, and $c_0$, $c_1$ are fitted constants (all provided in the SM). This formulation retains the key anisotropic features, systematically strengthening with increasing activity and area fraction. The \emph{central panel} of Fig.~\ref{fig:grthetafig} offers an additional validation of the analytical prediction, extracting radial cuts along the propulsion direction $\theta = 180^\circ$ (see SM for $\theta = 0^\circ, 90^\circ, 270^\circ$). These cuts show that the DNN, as well as Eq.~(\ref{eq:abp_angle_symbolic}), capture the near-contact peak and oscillatory behavior with remarkable accuracy. 

\paragraph{Conclusions.} 

We introduced a simple and generic method that combines particle-resolved simulations, deep neural networks (DNNs), and symbolic regression to predict microstructure in terms of analytical closed-form expressions. 
Beyond providing an efficient surrogate for simulations, our work paves the road towards structure-informed nonequilibrium theory. 
The generic character of the presented method invites a broad range of applications, e.g., to active systems with short-range attractions~\cite{sammullerDeterminingChemicalPotential2025}, in external potentials, and in confinement~\cite{nygard_anisotropic_2012}, as well as to sheared glassy and granular materials~\cite{fuchs_theory_2002,kranz_rheology_2018,dangeloRheologicalRegimesAgitated2025}. Finally, the closed-form expressions could inform novel inverse design strategies~\cite{leeMachineLearningbasedInverse2023,wangInverseDesignGlass2021}, and motivate a new wave of developments to predict dynamical properties directly from structural information in non-equilibrium systems~\cite{strickerUnifyingAtomsColloids2024,svetlizkySpatialCrossoverFarFromEquilibrium2021}.

\section{Acknowledgments}
B.L. and A.K.M. acknowledge funding by the Deutsche Forschungsgemeinschaft (DFG, German Research Foundation) in the framework of the collaborative research center Multiscale Simulation Methods for Soft-Matter Systems (TRR 146) under Project No. 233630050.

\bibliography{references}

\clearpage
\newpage

\setcounter{equation}{0}
\setcounter{figure}{0}
\setcounter{table}{0}
\setcounter{page}{1}

\renewcommand{\thefigure}{S\arabic{figure}} 
\renewcommand{\theequation}{S\arabic{equation}}

\onecolumngrid
\begin{center}
	\textbf{\large Supplemental Material: Learning microstructure in active matter}\\[.4cm]
	Writu Dasgupta,$^1$ Suvendu Mandal,$^1$ Aritra K. Mukhopadhyay,$^1$ and Benno Liebchen$^1$\\\vspace{0.2cm}
	\small $^1$ Institute for Condensed Matter Physics, Technische Universit{\"a}t Darmstadt, Hochschulstraße 8, 64289 Darmstadt, Germany.\\
\end{center}

\twocolumngrid

\section{Deep Learning Framework}

\subsection{Data Preprocessing and Architecture}
\begin{figure*}[tp]
	\centering
	\includegraphics[width=\linewidth]{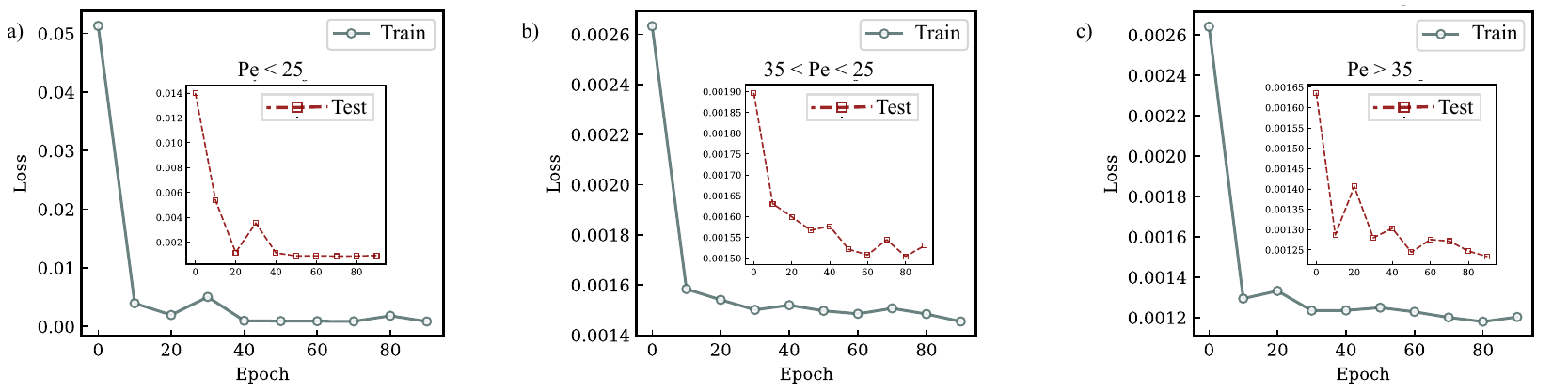}
	\caption{\textbf{Loss curves in curriculum learning:} Training and testing (inset) loss plotted against epoch number for (a) low ($\mathrm{Pe} \le 25$), (b) intermediate ($25 < \mathrm{Pe} \le 35$), and (c) high activity ($\mathrm{Pe} > 35$).}
	\label{fig:loss_abp_sym}
\end{figure*}

To predict the pair correlation function $\mathrm{g}$, we employ a feed-forward deep neural network (DNN). The complexity of the input space varies across the three studied regimes: (i) passive Brownian particles (inputs: packing fraction $\varphi$, distance $r$), (ii) isotropic active Brownian particles (inputs: P\'eclet number $\mathrm{Pe}$, $\varphi$, $r$), and (iii) anisotropic active Brownian particles (inputs: $\mathrm{Pe}$, $\varphi$, $r$, relative angle $\theta$).

As discussed in the main text, We generate a dataset using LAMMPS simulations ~\cite{LAMMPS} for $\varphi \in [0.20,0.50]$ and $\mathrm{Pe}\in[5,45]$. For each $(\varphi,\mathrm{Pe})$, we extract 20 statistically independent snapshots from our simulations. From these configurations, we compute $\mathrm{g}(r,\theta)$ with radial and angular resolution of $\Delta r=0.025$ and $\Delta\theta=4^\circ$. We normalize the simulation data to ensure numerical stability during training. The input features are scaled as follows: the distance $r$ and packing fraction $\varphi$ are used directly, as they naturally fall within sufficiently localized ranges ($r/\sigma \in [0, 5]$, $\varphi \in [0, 1]$). The P\'eclet number, which varies comparatively strongly ($\mathrm{Pe} \in [5, 45]$), is normalized via Min-Max scaling ~\cite{scikit-learn} to the range $[0, 1]$. The target variable $\mathrm{g}$ is log-transformed. This transformation prevents the high-magnitude values associated with the first coordination shell (where $\mathrm{g}(r)$ can exceed 20) from disproportionately dominating the loss function gradient and provoking instabilities.

The network architecture consists of three (for passive and Isotropic active Brownian system) or four (for anisotropic active Brownian system) fully connected hidden layers, each containing 256 neurons, and utilizes ReLU (only on the input layer) and LeakyReLU (applied on the hidden layers to get rid of the vanishing gradient problem). We utilize the AdamW optimizer ~\cite{loshchilov2017adamw} with a learning rate of $5 \times 10^{-4}$ (that progressively decays in case of curriculum learning for anisotropic system) and a weight decay of $1 \times 10^{-4}$. Extensive hyperparameter optimization confirms that this configuration provides an excellent balance between model expressivity and generalization capabilities.

\subsection{Loss Functions and Curriculum Learning}
We tailor the loss function to the physical complexity of the system. For passive and angle-averaged active systems, we minimize the Mean Squared Error (MSE). However, the anisotropic case introduces significant non-linearity and outliers due to the explicit $\theta$-dependence. To mitigate this, we employ a Smooth L1 Loss (Huber loss~\cite{huberloss}) function that handles those few outliers quite well without skewing the model disproportionately.

To accelerate convergence and avoid local minima, we implement a curriculum learning strategy ~\cite{10.1145/1553374.1553380}. We partition the training data into three regimes based on activity: low activity ($\mathrm{Pe} \le 25$), intermediate activity ($25 < \mathrm{Pe} \le 35$), and high activity ($\mathrm{Pe} > 35$) (see fig:~\ref{fig:loss_abp_sym}). The model is trained sequentially on these subsets for 100 epochs, progressively reducing the learning rate as the complexity of the input regime increases.

\section{Symbolic Regression Implementation}

We utilize the PySR software package ~\cite{cranmer2023interpretablemachinelearningscience} to discover analytical closed-form expressions that describe the DNN-generated surrogates. To ensure computational tractability, we do not train on the raw simulation data but rather on the smoothed predictions of the DNN, as specified in the following. 

\subsection{Dataset Selection and Subsampling}
While the passive and isotropic active cases allow for manageable dataset sizes ($< 10^4$ data points), the anisotropic case requires careful subsampling from the DNN predictions to avoid excessive computational costs. We generate a synthetic dataset covering the relevant parameter space ($\varphi, \mathrm{Pe}$) with radial cutoffs extending to $3\sigma$ to capture the second coordination shell.

To reduce the dataset from $10^5$ to a target of $10^4$ points while preserving structural details, we employ an importance-weighted subsampling strategy that prioritizes peak regions. We subsequently apply $K$-means clustering~\cite{K-means} to select representative points from the subsampled distribution.

\subsection{Model Configuration}
The symbolic regression evolves over $2 \times 10^4$ iterations of $20$ different population samples. We constrain the search space by limiting the maximum equation complexity to approximately $50$ operations and a tree depth of $8$. The operator pool includes standard algebraic functions, exponentials, and trigonometric functions, with a constraint preventing nested calls of the same operator (e.g., $\sin(\sin(x))$).
\begin{figure}[tp]
    \centering
    \includegraphics[width=\linewidth]{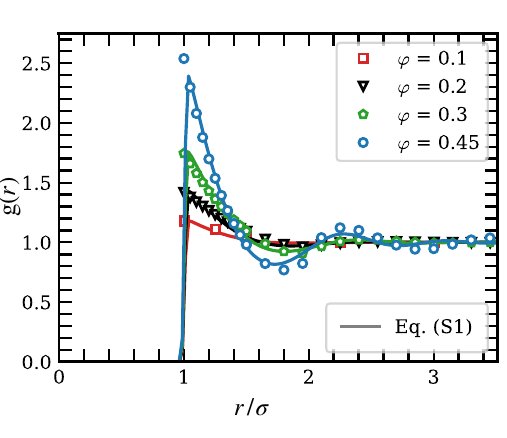}
    \caption{\textbf{Equilibrium radial distribution function for various packing fractions.} Symbols represent the 2D Percus-Yevick (PY) solutions, while solid lines correspond to the analytical expression derived through symbolic regression Eq.  (\ref{eq:pbp_lowphi}) for each packing fraction. }
    \label{fig:PBP_low_phi}
\end{figure}
The optimization objective is the minimization of the MSE between the candidate expression and the DNN predictions. If the algorithm fails to converge to a satisfactory expression within the iteration limit, we utilize a `warm start' procedure, re-initializing the search with the parameters of the best-performing equations from the previous run.

\section{Evaluation Metrics}

To provide a robust assessment of model performance, we report three complementary error metrics ~\cite{evaluation} . 
Below, we evaluate all three metrics for a given set of $N$ simulation reference values $\{y_i\}_{i=1}^N$ and model predictions $\{\hat{y}_i\}_{i=1}^N$, obtained either from the DNNs or from PySR:

\begin{enumerate}
    \item \textbf{Mean Absolute Error (MAE):}
    \[
    \mathrm{MAE} = \frac{1}{N}\sum_{i=1}^{N} |y_i - \hat{y}_i|
    \]
    The MAE quantifies the average magnitude of the error. It provides an intuitive measure of the typical discrepancy and is less sensitive to outliers than quadratic metrics.
    
    \item \textbf{Root Mean Square Error (RMSE):}
    \[
    \mathrm{RMSE} = \sqrt{\frac{1}{N}\sum_{i=1}^{N} (y_i - \hat{y}_i)^2}
    \]
    The RMSE penalizes large deviations heavily. This metric is particularly critical for assessing performance near the first coordination shell, where structural peaks are sharp and difficult to capture.
    
    \item \textbf{Coefficient of Determination ($R^2$):}
    \[
    R^2 = 1 - \frac{\sum_{i=1}^{N} (y_i - \hat{y}_i)^2}{\sum_{i=1}^{N} (y_i - \bar{y})^2}
    \]
    where $\bar{y}$ is the mean of the reference data. The $R^2$ score measures the fraction of variance captured by the model. A value near 1 indicates that the model reproduces both the mean behavior and the structural fluctuations of the pair correlation function.
\end{enumerate}

\begin{figure*}[tp]
	\centering
	\includegraphics[width=\linewidth]{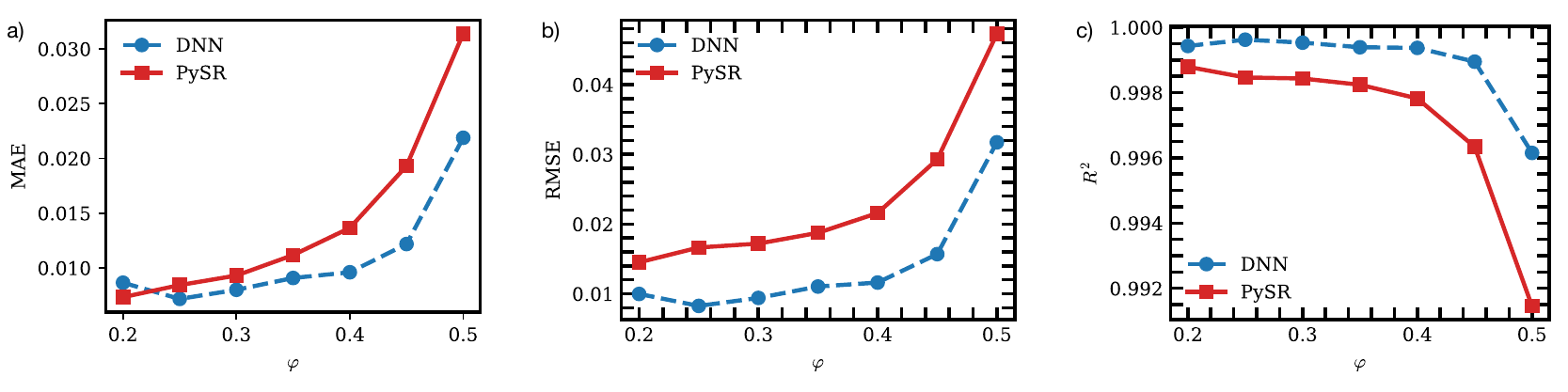}
	\caption{\textbf{Evaluation metrics for the passive Brownian system.} The metrics are plotted as a function of packing fraction $\varphi$. (a) Mean Absolute Error (MAE) and (b) Root Mean Square Error (RMSE) remain of order $\mathcal{O}(10^{-2})$, indicating high accuracy. (c) The coefficient of determination $R^2$ remains near unity. Blue circles denote DNN predictions; red squares denote symbolic regression results.}
	\label{fig:PBP_metrics}
\end{figure*}

Taken together, these three metrics provide a comprehensive characterization of predictive performance: MAE reflects typical absolute accuracy, RMSE highlights sensitivity to large localized errors, and $R^2$ quantifies how well the overall structure and variance of the data are 
captured. 

\section{Passive Brownian Particles}
The symbolic regression yields the following closed-form expression, valid for $\varphi \in [0.2, 0.5]$:

\begin{align}
    \mathrm{g}(r) &= \exp\left( \frac{(0.296\varphi)^r \, \sin(r^{2}(\varphi + 1))}{0.141} \right) \nonumber \\
    &\quad \times \cos^{48.933}\left[ \exp\left( -1.568r^{26.302} \right) \right].
\label{eq:pbp_lowphi}
\end{align}

To evaluate the validity of this analytical expression, we compare it with the 2D PY solutions (see Fig.~\ref{fig:PBP_low_phi}). The closed-form expression demonstrates good agreement with the 2D PY solutions.

Figure \ref{fig:PBP_metrics} illustrates the performance metrics for the passive case. As expected, we find that the DNN consistently achieves lower prediction errors (MAE and RMSE $\sim \mathcal{O}(10^{-2})$) compared to symbolic regression, but with a remarkably small performance gap. This gap reflects the trade-off between the high expressive capacity of the DNN and the interpretability constraint of the symbolic model. While the symbolic model is quantitatively less precise, it successfully captures the dominant structural features—specifically the periodicity and decay of the coordination shells. For both models, accuracy degrades moderately as $\varphi$ approaches 0.5.

\section{Active Brownian Particles (Isotropic)}

For the angle-averaged active case, the inclusion of activity leads to the following analytical expression:

\begin{equation}
    \mathrm{g}(r) = \sqrt{\exp\left[ \mathcal{A}(r,\varphi,\mathrm{Pe})\, \mathcal{B}(r,\varphi,\mathrm{Pe}) \right]},
\label{eq:abp_eqn_total}
\end{equation}
where the auxiliary functions are given as:
\begin{align}
    \mathcal{A}(r,\varphi,\mathrm{Pe}) &= -1.217\, r^{1-r}(\mathrm{Pe}\,\varphi)^{0.164} \nonumber \\
    &\quad + \frac{1.533(27.690 - \varphi)(r^{-29.139r})}{r}, \\[8pt]
    \mathcal{B}(r,\varphi,\mathrm{Pe}) &= \sqrt{\varphi} - \frac{64.653}{r^{24.784}} - \frac{1.265}{r} \nonumber \\
    &\quad - 0.989^{-\mathrm{Pe}}\,\varphi r \cos(r^{2.223}).
\end{align}

\begin{figure*}[tp]
	\centering
	\includegraphics[width=\linewidth]{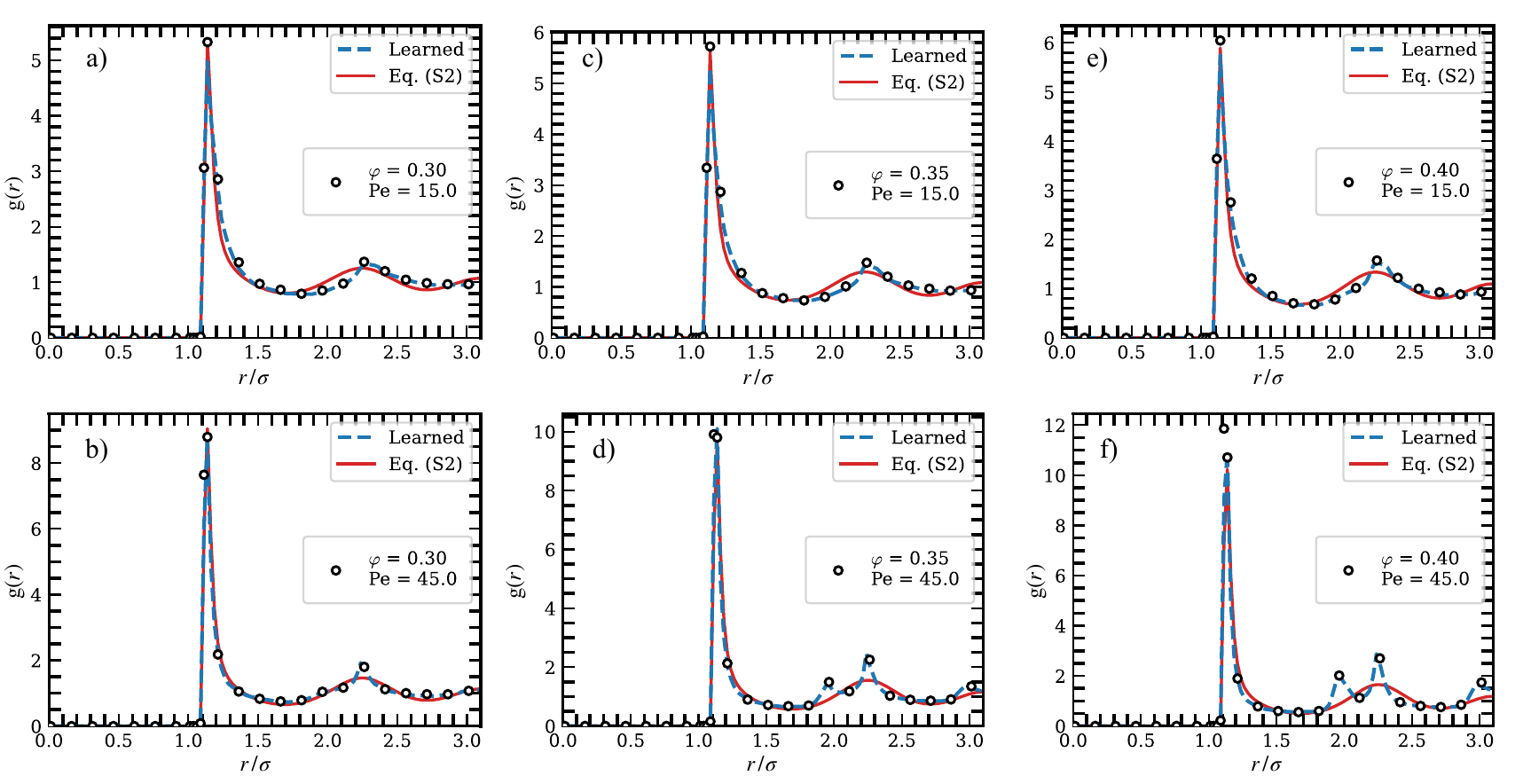}
	\caption{\textbf{Comparison of model predictions for active Brownian particles with isotropy.} The pair correlation function $\mathrm{g}(r)$ is plotted against normalized distance $r/\sigma$ for $\mathrm{Pe}=15$ (a,c,e) and $\mathrm{Pe}=45$ (b,d,f) at varying packing fractions. The symbol represents simulation data, blue dashed line denotes the learned (DNN) model prediction, and red solid line represents analytical expression given by Eq.~\ref{eq:abp_eqn_total}, derived through symbolic regression.}
	\label{fig:DNN_abp_sym}
\end{figure*}

Figure \ref{fig:DNN_abp_sym} compares the learned (DNN) model predictions against simulation data for varying P\'eclet numbers, demonstrating that the symbolic expression captures the shift in peak heights induced by activity.

\begin{figure*}[tp]
	\centering
	\includegraphics[width=\linewidth]{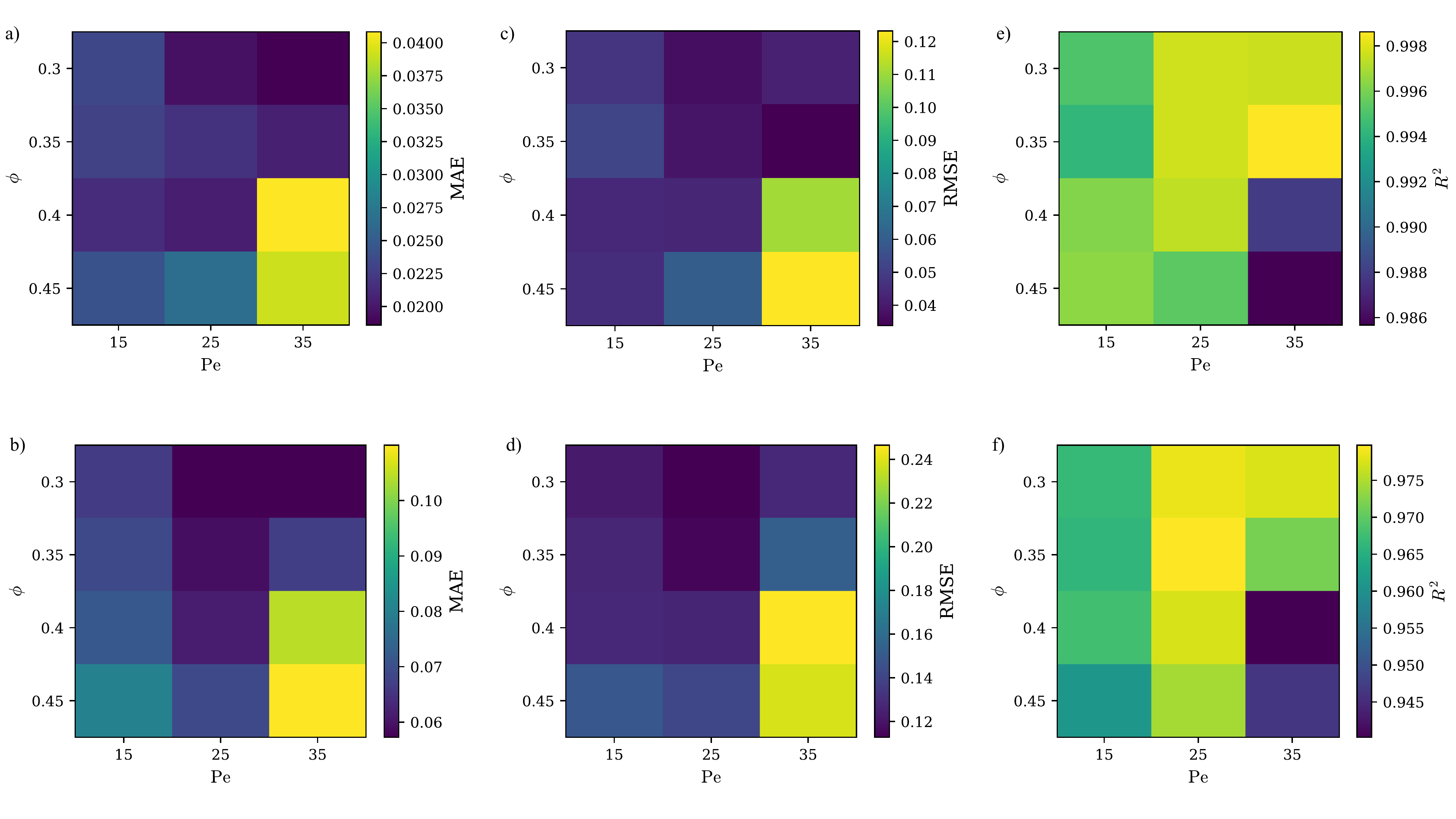}
	\caption{\textbf{Performance heatmaps for the isotropic active system.} Evaluation metrics are shown as a function of Péclet number $\mathrm{Pe}$ and packing fraction $\varphi$. Panels (a, c, e) display the MAE, RMSE, and $R^2$ for the DNN model, while (b, d, f) show the corresponding metrics for symbolic regression. Color intensity scales with the magnitude of the error.}
	\label{fig:eval_abp_sym}
\end{figure*}

We observe that errors increase systematically with both activity and density (see Fig. \ref{fig:eval_abp_sym}). The largest deviations occur in the high-activity, intermediate-density regime $(\mathrm{Pe} \approx 35, \varphi \approx 0.45)$, where motility-induced clustering creates sharp structural features that are challenging for the symbolic regression to capture fully. Nevertheless, the symbolic model retains qualitative fidelity also in this parameter regime.

\begin{figure*}[tp]
	\centering
	\includegraphics[width=\linewidth]{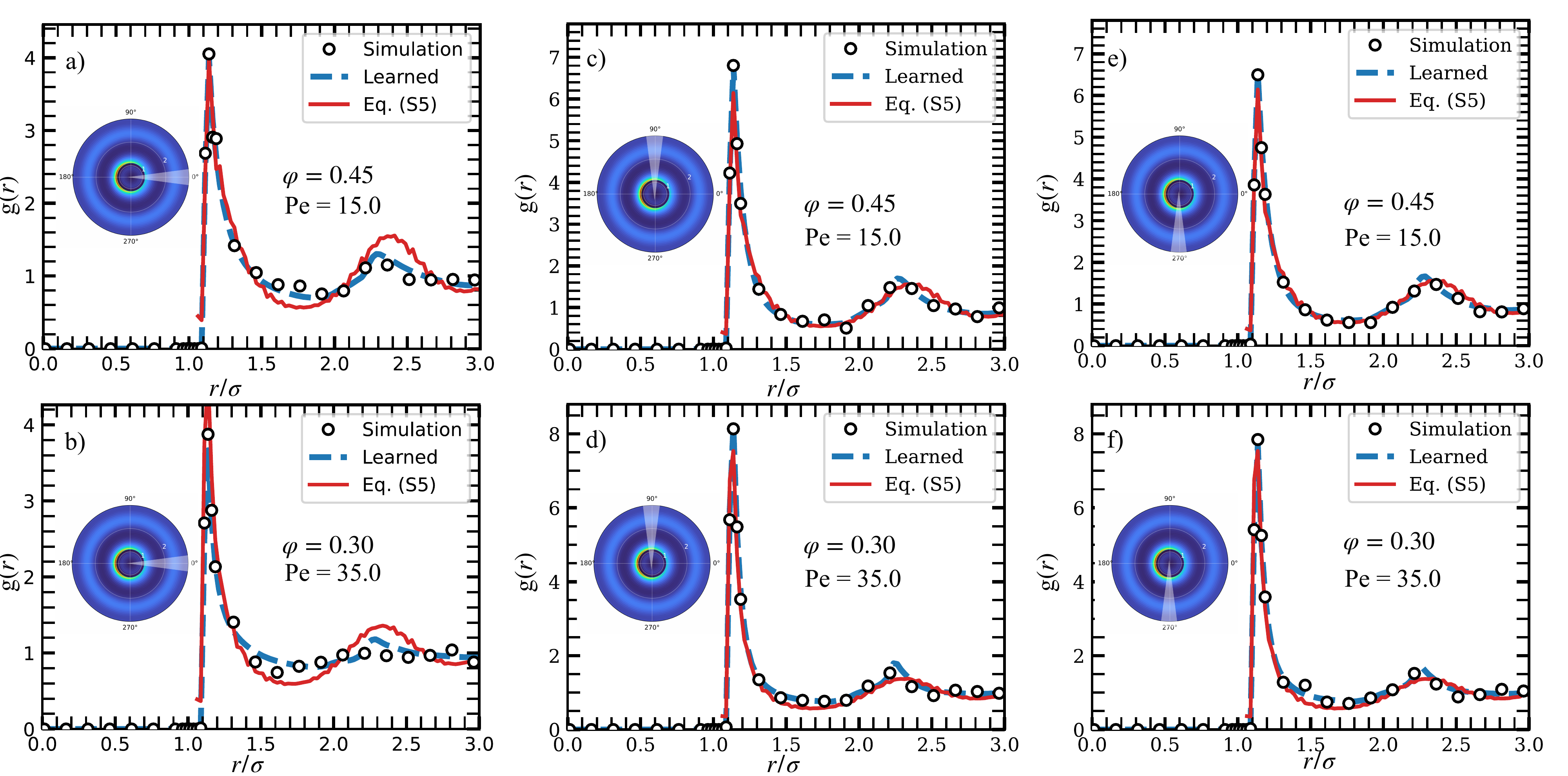}
	\caption{\textbf{Anisotropic pair correlation function cross-sections.} The radial dependence of $\mathrm{g}(r, \theta)$ is plotted along specific angular directions: (a, b) rear $\theta=0^{\circ}$, (c, d) lateral $\theta=90^{\circ}$, and (e, f) lateral $\theta=270^{\circ}$ ; (front $\theta=180^{\circ}$ is shown in the main text). Top panels correspond to $(\varphi, \mathrm{Pe}) = (0.45, 15)$, and bottom panels to $(\varphi, \mathrm{Pe}) = (0.30, 35)$. Black circles: simulation; blue dashed line: DNN; red line: symbolic regression. The symmetry between $90^{\circ}$ and $270^{\circ}$ confirms the physical consistency of the learned models.}
	\label{fig:cross_all}
\end{figure*}

\begin{figure*}[tp]
	\centering
	\includegraphics[width=\linewidth]{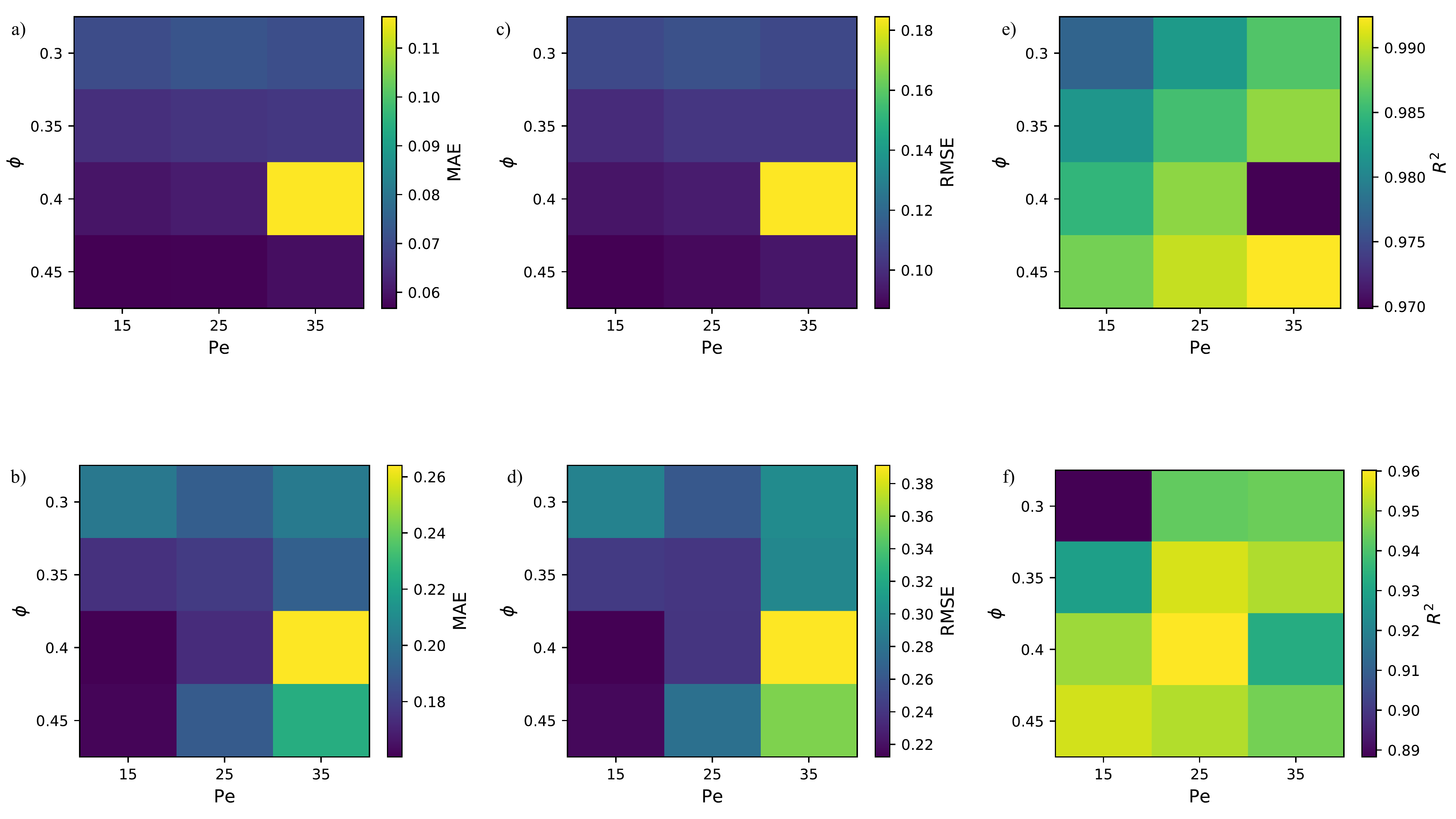}
	\caption{\textbf{Performance heatmaps for the anisotropic active system.} Evaluation metrics for $\mathrm{g}(r, \theta)$ are plotted against $\mathrm{Pe}$ and $\varphi$. Panels (a, c, e) show the DNN performance, while (b, d, f) display the symbolic regression performance. Errors are generally higher than in the isotropic case due to the complexity of directional correlations.}
	\label{fig:eval_abp_asym}
\end{figure*}

\section{Anisotropic Pair Correlation Function}

The anisotropic pair correlation function $\mathrm{g}(r,\theta)$ represents the most complex scenario, requiring the model to resolve directional symmetry breaking. The symbolic regression identifies the following functional form:

\begin{align}
\mathrm{g}(r,\theta) & = r\,\exp\Biggl\{ \Biggl[ \varphi + \bigl(3.295 - \cos\theta\,(0.777-\varphi) \bigr) \nonumber\\
&\qquad\sin\!\left(\frac{0.071}{r-1.072}\right)\Biggr]\times \sin\!\Biggl[ \varphi + \mathrm{Pe^*} + \frac{r}{0.175} - \nonumber\\
&\qquad\frac{\cos\theta-\sin\!\left(\frac{-r}{0.010}\right)}{7.828} \Biggr]- 1.275\Biggr\} + 0.328 .
\end{align}
where, $\mathrm{Pe^*}$ is the (min-max) scaled P\'eclet number.
To validate the physical consistency of our models, we analyze cross-sections at varying angles (see Fig. \ref{fig:cross_all}). The front direction ($\theta = 180^{\circ}$) exhibits maximal particle accumulation due to persistent self-propulsion, while the rear ($\theta = 0^{\circ}$) shows depletion. Crucially, the comparison between the lateral directions $\theta = 90^{\circ}$ and $\theta = 270^{\circ}$ demonstrates that both the DNN and the symbolic equation respect the mirror symmetry of the system.
    
The evaluation metrics indicate that while error magnitudes are higher than in the isotropic cases, the DNN maintains high accuracy (Fig. \ref{fig:eval_abp_asym}). The symbolic regression captures the essential angular modulation and the dominant radial structure, offering a tractable analytical approximation for the highly non-linear anisotropic microstructure.

\end{document}